\documentclass[10pt,amsmath,amssymb,latexsym]{article}
\usepackage{graphicx}  
\usepackage{dcolumn}   
\usepackage{bm}        
\usepackage{verbatim}  
\usepackage{mathtools}  

\newcommand{\ket}[1]{\left|{#1}\right>}
\newcommand{\ketBasis}[2]{\left|{#1}\right>_{#2}}
\newcommand{\bra}[1]{\left<{#1}\right|}

\newcommand{\braBasis}[2]{\prescript{}{#2}{\left<{#1}\right|}}

\newcommand{\braketBasisLeft}[3]{\prescript{}{#3}{\left<{#1}|{#2}\right>}}
\newcommand{\Tr}[1]{\textrm{Tr}\left({#1}\right)}
\newcommand{\PartTr}[2]{\textrm{Tr}_{#1}\left({#2}\right)}
\newcommand{\toshIdentity}{{\mathbf{1}}}

\begin{document}

\title{The issue with the initial state in quantum mechanics}
\author{Hitoshi Inamori\\
\small\it Soci\'et\'e G\'en\'erale\\
\small\it Boulevard Franck Kupka, 92800 Puteaux, France
}

\bigskip

\date{\today}

\maketitle

\begin{abstract}
In the conventional formulation of quantum mechanics, the initial description is given only for the physical system under study. It factors out the state for the experimenter. We argue that such description is incomplete and can lead to statements which can in theory be meaningless. We propose that within a complete description, the initial state must include  the state of the experimenter. With such formulation quantum mechanics provides joint probabilities for conjointly observed events, rather than a probability conditional on some initial state for the system under study. This feature is desirable, as with quantum mechanics, statements on what happened in the past may have no meaning in the present.

\bigskip

\end{abstract}

\section{The initial state in the conventional formulation of quantum mechanics}  

In quantum mechanics, a physical system $B$ is described by a vector or ``state'' in a Hilbert space $H_B$. In a conventional quantum mechanical description of an experiment, we say that the system $B$ is initially in a normalized state $\ket{\chi_a}$ at a time $t_A$, with probability $p(a)$, where $a=1,2,\ldots$ correspond to possible initial states.

In the absence of a measurement, any evolution which is physically allowed is described by a unitary operator acting on the Hilbert space for the system under study. In this case, let's denote by $V_B$ the unitary operator acting on $H_B$, transforming the initial state $\ket{\chi_a}$ at time $t_A$ into the state $V_B\ket{\chi_a}$ at a later time $t_B$.  Suppose that a measurement is performed on $B$ at $t_B$ after this evolution. A measurement can be described as a projective measurement onto an orthonormal basis $\{\ketBasis{j}{B}\}_{j}$ of $H_B$, and the conditional probability of obtaining the measurement outcome $b$ given that initial state was $a$ is given by the Born rule $p(b|a)=|\braBasis{b}{B} V_B \ket{\chi_a}|^2$. 
For ease of notation, we define the basis $\ketBasis{j}{B(t_A)}=V_B^\dagger\ketBasis{j}{B}$, in which case we obtain $p(b|a)=|\braketBasisLeft{b}{\chi_a}{B(t_A)}|^2$. The conventional formulation of quantum mechanics gives conditional probabilities of an event at $t_B$, conditional on a realized preparation at the beginning of the experiment $t_A$. 
This is the description of quantum mechanical laws that is usually taught in physics textbooks.

\section{A complete description of the initial state in quantum mechanics}
The purpose of this note is to show that this conventional description does not give a complete picture for the initial state of the experiment.

Following the well known insight by Everett~\cite{Everett}, we need to understand that the initial state for the system $B$ above is the result of a prior physical process involving the experimenter. Let's call $A$ the experimenter, and denote by $H_A$ the Hilbert space associated with $A$. When we say that the system $B$ is in a state $\ket{\chi_a}$, what really happened, without loss of generality, is that the experimenter $A$ interacted with $B$ leading to a superposition of states
\begin{equation}
\ket{\psi}=\sum_{i}\alpha_i\ketBasis{i}{A}\otimes \ket{\chi_i}\label{superposition}
\end{equation}
and happens to observe the outcome $a$. Here $\{\ketBasis{i}{A}\}_{i=1,2,\ldots}$ is a set of some orthonormal states in $H_A$, where each $\ketBasis{i}{A}$ corresponds to a situation in which $A$ observes the outcome $i$ for the initial preparation.

Now, as discussed in~\cite{Everett}, the system made of $A$ and $B$, denoted by $A\otimes B$, is an isolated quantum system seen from the outside, and as such should follow a unitary transformation.  

Therefore, the initial state of the system after the preparation of $B$ remains a superposition of states in which each possible outcome for the measurement by $A$ is present. The complete initial state for $A\otimes B$ is not a classical probabilistic mixture of states as described in the conventional description of quantum mechanics, but a superposition of joint states in $A\otimes B$ in which states for $A$ and states for $B$ are entangled. 

In other words, the initial state for the experiment should not be a density operator $\rho= \sum_a p(a) \ket{\chi_a}\bra{\chi_a}$ describing a classical mixture in which $B$ is in state $\ket{\chi_a}$ with probability $p(a)$, but a superposition of states in $A\otimes B$ as represented in Equation~(\ref{superposition}).

With this complete description for the initial state, quantum mechanics does no longer provide a conditional probability based on a supposed realization of an initial preparation. Rather, quantum mechanics provides joint probabilities for the observation of $a$ and the outcome $b$ of the experiment.

As the set $\{\ketBasis{j}{B(t_A)}\}_j$ form a basis for $H_B$, the states $\ket{\chi_i}$ can be written as $\ket{\chi_i}=\sum_j \beta_{i j} \ketBasis{j}{B(t_A)}$ for some set of complex numbers $\beta_{i j}$ and therefore the initial state reads as:
\begin{equation}
\ket{\psi}=\sum_{i j}\mu_{i j}\ketBasis{i}{A}\otimes\ketBasis{j}{B(t_A) }\label{superposition2}
\end{equation}
where $\mu_{i j} = \alpha_i \beta_{i j}$. Because the states $\ket{\chi_i}$ are normalized, we have  $\sum_{j}|\mu_{i j}|^2=|\alpha_i|^2\sum_j |\beta_{i j}|^2=|\alpha_i|^2$.

The joint probability of having the initial preparation $a$ and the experiment outcome $b$ is now
\begin{eqnarray}
p(a,b)&=&\Tr{\toshIdentity_A\otimes V_B \ket{\psi}\bra{\psi} \toshIdentity_A\otimes V_B^\dagger
	\ketBasis{a}{A}\otimes\ketBasis{b}{B} \braBasis{a}{A}\otimes\braBasis{b}{B}} \\
&=& |\mu_{a b}|^2
\end{eqnarray}
using the relation $\ketBasis{j}{B(t_A)}=V_B^\dagger\ketBasis{j}{B}$.

\section{Prediction and Retrodiction}
Because the complete description of the initial state leads to joint probabilities, it is straightforward to obtain marginal and conditional probabilities:
\begin{eqnarray}
p(A=a)&=& \sum_j p(A=a,B=j) = \sum_j |\mu_{a j}|^2\\
p(B=b|A=a)&=& \frac{p(A=a,B=b)}{p(A=a)} = \frac{|\mu_{a b}|^2}{\sum_j |\mu_{a j}|^2}\\
p(B=b)&=& \sum_i p(A=i,B=b) = \sum_i |\mu_{i b}|^2\label{pb}\\
p(A=a|B=b)&=& \frac{p(A=a, B=b)}{p(B=b)} = \frac{|\mu_{a b}|^2}{\sum_i |\mu_{i b}|^2}
\end{eqnarray}
where for the sake of clarity, we have added the system being observed in the notation.

Note that the formula which gives the conditional probability for the final event $b$ based on the initial event $a$, $p(B=b|A=a)$, is symmetric with the formula giving the conditional probability for the initial event $a$ based on the final event $b$, $p(A=a|B=b)$.

The conditional probability $p(B=b|A=a)$ is a predictive one. Given the initial state $a$, we try to guess what the final state $b$ is going to be at a later time. This is the conditional probability that is provided by the conventional formulation of quantum mechanics, and which is of practical interest most of the time.

In comparison, the calculation of the conditional probability $p(A=a|B=b)$ is a retrodictive one. We try to guess what was a past state $a$ from the realization of a final state $b$. 

There is an interesting literature~\cite{Watanabe, Pegg, Aharonov}  about the way one can try to retrodict past event probabilities from the observation of the present state. Many papers note that an asymmetry appears between prediction and retrodiction in the following sense: on the prediction side, once the realization $a$ for the initial state is known, one can deduce the conditional probability distribution for the final outcome $b$ without any further assumption, as given by the Born rule:
\begin{equation}
p(B=b|A=a)= \Tr{\ketBasis{b}{B}\braBasis{b}{B}V_B\ket{\chi_a}\bra{\chi_a}V_B^\dagger}=|\beta_{a b}|^2.
\end{equation} 

This should come as no surprise as the predictive conditional probability is precisely what is given directly by the conventional formulation of quantum mechanics.

 By contrast, the last formula does not allow to compute the conditional probability for $a$ given a realization of $b$. This is to be expected as one cannot deduce conditional probabilities $p(A=a|B=b)$ when one is only provided with the conditional probabilities $p(B=b|A=a)$, because
 
 \begin{eqnarray}
 p(A=a|B=b)&=&p(B=b|A=a)\frac{p(A=a)}{p(B=b)}\\
 &=&p(B=b|A=a)\frac{p(A=a)}{\sum_{i} p(B=b|A=i)p(A=i)},
 \end{eqnarray}
 and that the probabilities $p(A=i)$ are not given by the above formula. It is true that one can add the information about the marginal probabilities for $a$ by expressing the initial state as a classical mixture of states $\rho=\sum_a p(A=a)\ket{\chi_a}\bra{\chi_a}$, however the predictive formula and the retrodictive formula still bear a mathematical asymmetry in the way they are deduced in the conventional description of quantum mechanics. 
 
 In the next section we will identify the origin of this asymmetry with an effort to give more physical insight. We propose that the correct description of a quantum state should include the state of the experimenter. The conventional formulation in which the experimenter is factored out is -- as far as theory is concerned -- a special case which is correct only under certain conditions.

\section{The trouble with the conventional formulation of quantum mechanics}

Let's come back to the complete description of the initial state as given in Equation~(\ref{superposition}). A partial description of the state restricted to the system $B$ is obtained by tracing over the Hilbert space $H_A$:

\begin{equation}
\rho_B=\PartTr{A}{\ket{\psi}\bra{\psi}}= \sum_i |\alpha_i|^2 \ket{\chi_i}\bra{\chi_i}= \sum_{i} p(A=i) \ket{\chi_i}\bra{\chi_i}
\end{equation}
as $p(A=i)=\sum_j|\mu_{i j}|^2=|\alpha_i|^2$. Without surprise, we find the mixture of states for $B$ with which we would start in the conventional formulation of quantum mechanics. Therefore, provided that the experiment only affects the system $B$, the complete description presented in this note gives results that are identical to what the conventional description gives.
 
 What is the point of introducing the complete description if it gives exactly the same results as the conventional one?
 
 Note that in the reasoning above, we had to assume that the experiment was only affecting the system $B$, that is, the experiment was not affecting the system $A$. In particular, we had to assume that after the preparation of the initial state, the system $A$  was no longer interacting with the system $B$. One could defend such hypothesis by arguing that the result $a$, being witnessed by an experimenter, is a classical information encoded in a macroscopic state that cannot be altered by the experiment. 
 
 But is this really the case?
 
 The result $a$, as any information, must be stored in a physical system. The outcome $a$ has no longer meaning if this physical system is altered. The statement ``$B=b$ at time $t_B$, conditional on $A=a$ at time $t_A$" has a meaning only if at the time when the statement is made, we can observe a physical proof that $A$ was $a$ at the past time $t_A$.
 
 Suppose that $A$ shows the outcome $a$ at time $t_B$. Does this prove that $A=a$ at time $t_A$? We could argue that the system $A$ is an isolated system after the preparation of $B$ and that $A$ could not evolve between times $t_A$ and $t_B$ or that $A$ is macroscopic enough not to evolve between $t_A$ and $t_B$. 
 
 However it seems more likely that $B$ interacts in one way or another with the system $A$ even after the initial preparation of $B$. Indeed $A$ had interacted with $B$ at some prior time for the preparation of $B$, and it seems difficult to accept that all possible interactions can been turned off perfectly between $A$ and $B$ after $t_A$. Therefore, at least in theory, the fact $A=a$ at time $t_B$ does not guarantee that the statement $A=a$ at time $t_A$ is true. As a consequence, the statement ``$B=b$ at $t=t_B$ conditional on $A=a$ at $t=t_A$'' has no rigorous meaning in theory, because we cannot ascertain that $A=a$ at time $t=t_A$ when $B$ is observed. The only fact that can be ascertained when $B$ is observed, ultimately, is what is observed conjointly with the observation of $B$. In our case, we cannot ascertain that $A=a$ at time $t=t_A$, but we can ascertain that $A=a$ at time $t=t_B$ if $A$ is observed conjointly with $B$.

  We can make statement only about relationships between observations made simultaneously or conjointly. This is precisely what the complete description of initial state gives: indeed, if interaction between $A$ and $B$ cannot be ruled out after $t_A$, nothing prevents us from describing the interaction, say $U_{A B}$, between $A$ and $B$ after $t_A$. This is possible because we have kept a complete quantum description for $A\otimes B$:
  \begin{equation}
  \ket{\psi}\mapsto U_{A B}\ket{\psi}
  \end{equation}
  and then we can compute the joint probability for the outcomes obtained from the conjoint observation of $A$ and $B$ at time $t_B$:
  \begin{equation}
  p(A=a,B=b)=\Tr{U_{A B} \ket{\psi}\bra{\psi} U_{A B}^\dagger
  	\ketBasis{a}{A}\otimes\ketBasis{b}{B} \braBasis{a}{A}\otimes\braBasis{b}{B}}.
  \end{equation}

  The complete description gives joint probabilities of events that are observed at a same time, in a conjoint observation. It does not attempt to give a relationship with a past event that is no longer observable, but a relationship between two simultaneous events that are observable. In this sense, the complete description adopts a fundamentally different view from the conventional one: it abandons the notion of a realized event in the past. Instead, it gives relationship between direct observations that are done conjointly or simultaneously. 
 
 \section{Conclusion}
 
 We have argued that a complete description of an initial state must encompass the experimenter who has entangled himself or herself with the physical system under study. By reducing the initial state to a mixture of states describing the system under study only, the conventional formulation of quantum mechanics neglects potential interaction between the studied system and the experimenter after the preparation phase. Such approximation may well be justified in practice, but in theory there seem to be no way to guarantee the independence of the experimenter and the system under study during the experiment. Taking the initial state for the complete joint system including the experimenter allows to circumvent this issue properly, within a rigorous and unambiguous formalism.

 Past event does not have an existence per se and must be encoded in a physical system.
 By accepting that we cannot neglect the interaction between the physical system encoding the outcome of the preparation and the evolution of the physical system under study, we acknowledged that past event may not have an unambiguous definition at the end of the experiment. The set of observables that do have unambiguous meaning taken together are observables that are witnessed conjointly. Describing the quantum system for the complete system including the experimenter allows to give joint probabilities for such set of observables. The conventional formulation, by introducing the notion of past event, cannot be always consistent because the past event may not even be defined by the time the experiment's outcome is observed.
 
 Obviously, we can introduce physical system that do serve as a ``marker'' for the past, such as the experimenter memory or a written note describing the outcome of the preparation phase. These systems can be included in the complete description of the initial state. They can be observed at the end of the experiment, and we could interpret this outcome as reflecting events that happened in what we call the ``past''~\cite{Inamori}. However, these markers can themselves interact with the system under study, and as such cannot serve as a proof of the events in the past, or even demonstrate that the past in which such event happened existed at all.

We commonly assume that what we witness in the present is explained -- at least partially -- by what happened in the past. Physical laws describe the relationship between such past events and present observations. However, everything we know is ultimately based on present observation: we may categorize some data, such as the output from our memory or the writings in a notebook, as coming from a ``past'', but such categorization remains observer-dependent.

Classical physics allows the existence of a past that has an unambiguous definition and that is completely deterministic. This is because classical physics itself is fully deterministic. There is only one trajectory allowed for the state of the system once it is completely known at some instant. As such, assuming the existence of a deterministic past does not lead to any inconsistency.

Quantum mechanics does not allow such certainty, and as we have seen in this note, attempting to introduce a notion of deterministic past condition does introduce inconsistencies under general situations. In quantum mechanics, the only known data are the data obtained conjointly from a single measurement. Physical laws do no longer provide relationship between some past and present. Rather they provide relationship between two categories of data, the one which the observer classifies as ``coming from the past" and the other which the observer classifies as ``coming from the present".  Both sets of data are nevertheless obtained conjointly from a single measurement. 
  

\end{document}